# A heuristic resolution of the Abraham–Minkowski controversy

**Guoxu Feng[a], Jun Huang[b]**

School of Aeronautic Science and Engineering, Beihang University, Beijing, 100191, China

[a] e-mail: fengguoxu007@126.com (corresponding author)

[b] e-mail: junh@china.com

**Abstract** This paper reviews the history and origin of the Abraham–Minkowski controversy and points out that it is a continuation of the controversy over the speed of light in medium. Upon considering an aircraft flying at a constant speed along the great–circle route from the perspective of a geosynchronous space station, we show that the A–M controversy arises from non–local observation. The relative motion refractive index and the gravitational field refractive index are defined by the space–time metric tensor, which reveals the A–M controversy hidden in special and general relativity. As another example, we take light propagating over the surface of the sun, and show that Minkowski and Abraham forces are responsible for the gravitational deflection of light and the Shapiro time delay, respectively. Overall, we heuristically conclude that non–local observation is the cause of the A–M controversy.

## 1 Introduction

What is the momentum of light in transparent medium such as water and glass? This seems to be a very simple question. Yet, no clear answer has been provided so far [1,2]. At the beginning of the 20th century, Minkowski and Abraham gave two opposite answers based on the theory of classical electrodynamics [3,4]. For more than a hundred years, both approaches have been supported by several experiments [5–27]. Some experiments seem to clearly prove that one or the other is correct, but the theoretical explanations of the experimental phenomena are quite different [28–38], and even two opposing viewpoints may be used to explain the same experiments [38]. There is a wide literature that deals with this topic, and the number of theoretical papers is about 25 times more numerous than those dealing with experiments [39]. Among these theories, some support Abraham [40–44], some support Minkowski [46–51], some believe that both are true and that they are different forms of momentum [52–56], and some believe that neither Abraham nor Minkowski is true [57–62]. This divergence and confusion about the momentum of light in a medium is usually referred to as the Abraham–Minkowski controversy.

Here, we first briefly review the experiments related to the A–M controversy, and point out that the fundamental reason for its appearance is the opposition of experimental observational evidence. It also links the A–M controversy to the controversy about the light speed in medium, proving that the former is a continuation of the latter. Then, taking an aircraft flying at a constant speed along the great–circle route as an example, the motion of the aircraft is analyzed from the perspective of geosynchronous space station (i.e. a non–local observer). It is proved that the equation expressing the conservation of angular momentum of the aircraft from the perspective of geosynchronous space station is equivalent to the equation of great–circle route from the perspective of earth. The A–M controversy for the aircraft is shown to arise from the non–local nature of the observations from the geosynchronous space station. Then, the relative motion refractive index and gravitational field refractive index are defined by using the metric of flat space–time and curved space–time respectively. The physical phenomena related to special relativity and general relativity are reexamined from the perspective of non–local observer, and the A–M controversy hidden in relativity is pointed out. Finally, by considering the light propagating in the gravitational field of the sun, we prove that the Minkowski and Abraham forces may be used to account for the gravitational deflection of light and the time delay of Shapiro, respectively. The overall conclusion of the paper is that the non–local observation of physical phenomena is the cause of the A–M controversy.

## 2 The history and origin of controversy

### 2.1 History

In 1908, Minkowski gave the first form of electromagnetic momentum density in a medium, $\vec{g}_M = \vec{D} \times \vec{B}$ [3]**.** In 1909, Abraham gave a second form of electromagnetic momentum density in a medium 1909, $\vec{g}_A = (\vec{E} \times \vec{H})/c^2$ [4]. Where, $\vec{D}$ is the electric displacement, $\vec{B}$ is the magnetic flux density, $\vec{E}$ is the electric field intensity, $\vec{H}$ is the magnetic field intensity, and $c$ is the light speed in vacuum. By volume integration of $\vec{g}_M$ and $\vec{g}_A$, Minkowski momentum $p_M$ and Abraham momentum $p_A$ are introduced as in Eqs. (1a) and (1b) [18,52]:

$$p_M = np_0 = nm_0 c = n\frac{h}{\lambda_0} = \frac{h}{\lambda} \tag{1a}$$

$$p_A = \frac{p_0}{n} = \frac{m_0 c}{n} = \frac{h}{n\lambda_0} = \frac{h}{n^2 \lambda} \tag{1b}$$

where, $n$ is the refractive index of the medium, $p_0$ is the vacuum momentum of photon, $m_0$ is the mass of the photon in vacuum, $h$ is the Planck constant, $\lambda_0$ is the wavelength of light in vacuum, and $\lambda$ is the wavelength of light in the medium.

In 1912, Barlow found that the oblique passage of a beam of light through a plate of refracting material produces on the matter of the plate a torque, whose magnitude may be deduced from the transfer of momentum in the beam. Some of the experimental results supported for Minkowski momentum within the experimental errors, but another part of the experimental results is characterized by errors too large to draw a definite conclusion [5]. In 1951, Jones reported an experiment involving the radiation pressure caused by light on a plane mirror immersed in a liquid [6]. Three years later, Jones and Richards discussed the measurement process and results in detail in another article, which supported Minkowski momentum within the margin of error of ±1.2% [7]. In 1978, Jones, F.R.S. and Leslie repeated the experiment 27 years ago replacing the mirror with a high reflectivity and low absorptivity one, and increased the accuracy of the experiment to 0.05%, thus more strongly supporting the Minkowski momentum [8]. Although the light pressure experiment of Jones et al. is considered as one of the most powerful proofs of Minkowski momentum [18], Gordon proposed a different theoretical explanation [28]. Webb even used the force derived from Abrahams momentum to explain the results of Jones's experiments [38].

In 1973, Ashkin and Dziedzic observed the radiation pressure on the free surface of a transparent liquid medium using a focused pulsed laser. They showed that light either entering or leaving the liquid exerts a net outward force at the liquid surface, the total deformation of the liquid surface being in the order of microns. The results support the Minkowski momentum [9]. In recent years, with the development of microstructural deformation measurement technology, together with numerical simulations, some scholars have measured the gas–liquid interface deformation caused by light pressure at the nanoscale and carried out numerical simulation based on theoretical models. The results are in good agreement with Minkowski momentum [10–13]. However, similar experiments conducted by Zhang and She et al. after changing the experimental parameters showed the sunken of the gas–liquid interface caused by light pressure, thus supporting the Abraham momentum [14]. Leonhardt explained the two opposite experimental results and proved that the deformation of interface caused by optical pressure is a hydrodynamic problem [54].

In 1980, Gibson et al. measured the photon drag effect in germanium and silicon, and the effects were in good agreement with the Minkowski momentum under low frequency conditions [15]. In 1994, Kristensen and Worrdman measured the angular momentum of light in a medium, and found that the angular momentum of photon is independent of the refractive index of the medium [16]. According to this result, the Minkowski momentum can be proved to be correct indirectly by using Bouguer's law [63]. In 2005, Campbell et al. have measured a systematic shift in the photon recoil frequency due to the index of refraction of the condensate, and the results support Minkowski momentum [17]. In 2011, Wang et al. indirectly proved that Minkowski momentum is correct through the reversed Fischer experiment [18]. In 2019, Schaberle et al. used photo acoustic detection of ultrasound waves generated when pulsed laser light meets the water/air interface at 3.9°C (zero thermal expansion), to distinguish momentum transfer from thermoelastic effects. Under the condition of total internal reflection, the results are consistent with Minkowski momentum [19].

In 1975, Walker et al. conducted accurate measurements of the Abraham density force in a $BaTiO_3$ ceramic using time–varying electric field and time invariant magnetic field. The measured torque amplitudes were in good agreement with the predicted values within an error range of 10% [20,21]. Two years later, Walker et al. measured the Abraham density force using time–invariant electric field and a time–varying magnetic field [22]. Two years after that, Lahoz and Graham measured the Abrahan force inside a low permittivity nonmagnetic material using low frequencies, and the corresponding experimental error were about 20% [23]. In 2008, She and Yu et al. reported a direct observation of the inward push force on the end face of a free nm fiber taper exerted by the outgoing light [24]. This experiment seems to be the simplest, and the most direct observation of Abraham momentum. However, the author's theoretical explanation of experimental phenomena has been questioned and opposed by other scholars [32,33].

In 2012, using a time–dependent electromagnetic field, Rikken and van Tiggelen observed the intrinsic Abraham force in a dielectric in contact with a conducting liquid [25]. In 2017, Choi et al. experimentally demonstrated the isolation and measurement of the Abraham force of an optical wave on a dielectric medium, by utilizing the adiabatic mode transformation along a liquid–filled hollow optical fiber. The authors also analyzed the Abraham force numerically, using finite element methods to determine the fundamental optical mode distributions, showing good agreement with experimental results [26]. In 2017, Kundu studied the bending deformation of Graphene Oxide film surface caused by low–power focused laser irradiation, which provided observational evidence for the Abrahan force [27]. However, Brevik did not agree with Kundu's explanation for the experimental phenomenon, and believed that the experiment could not be used as evidence of the existence of the Abraham force [35].

## 2.2 Origin

It is common sense that a light beam entering water does not change its color. The color is determined by the frequency $f$, which being a property of the source, remains unchanged after entering the medium. It can be seen from $E = hf = m_0 c^2$ that the assumption of constant frequency $f$ is equivalent to the assumption of constant mass $m_0$ of the photon. Upon writing, Minkowski and Abraham speeds $v_M$ and $v_A$ as shown in Eqs. (2a) and (2b), we see that the A–M controversy is equivalent to the controversy over the speed of light in medium, which dates back even further to the 17th century.

$$v_M = \frac{p_M}{m_0} = nc \tag{2a}$$

$$v_A = \frac{p_A}{m_0} = \frac{c}{n} \tag{2b}$$

In 1637, Descartes in his La Dioptrique gave the law of refraction of light as shown in Eq. (3) [64].

$$\frac{\sin\theta_1}{\sin\theta_2} = k = \frac{v_{M2}}{v_{M1}} \tag{3}$$

where, $\theta_1$ is the angle of incidence, $\theta_2$ the angle of refraction, and $k$ a constant that depends only on physical medium that the ray of light goes through, e.g., air and water. $v_{M1}$ and $v_{M2}$ are the speeds of light in the two medium, and they are equivalent to Minkowski speed.

Later, it was found that Snell gave the formula in his unpublished manuscript as early as 1621 [65]. So now, except for the French (Descartes is French, Snell is Dutch), most people call it Snell's law [65]. Descartes assumed that light is made up of small discrete particles called "corpuscles" (little particles) which travel in a straight line with a finite velocity and possess impetus. Based on this, the law of refraction was obtained [65]. Descartes also used the law to derive the hyperbolic form of perfect lenses that can focus incoming parallel rays to a single point [65].

However, Descartes' view has been criticized by Fermat, who believes that light propagation along the shortest path in time [64]. Around 1661, Fermat gave the refraction law as shown in Eq. (4) according to the principle of minimum time [64].

$$\frac{\sin\theta_1}{\sin\theta_2} = k = \frac{v_{A1}}{v_{A2}} \tag{4}$$

where, $v_{A1}$ and $v_{A2}$ are the speeds of light in two medium that Fermat thinks respectively, and they are equivalent to Abraham speed.

Although Fermat got the same refraction law as Descartes, that is, the ratio of the sine of the incident angle and the refraction angle is equal to the constant $k$, his view on the speed of light in the medium is completely opposite to that of Descartes.

Newton inherited and developed Descartes' view [66,67]. He thought of light as a stream of small particles. In passing from one medium to another of different density, the particles would be attracted to the denser one by a refractive force normal to the interface. The refractive force would instantaneously modify the square of a light particle's velocity by a fixed amount. If the motion were into the denser medium, the change would be an increase. The component of velocity parallel to the interface would not be affected; the increase would appear only in the perpendicular component. In his Mathematical principles of natural philosophy (1687), Newton proved that Snell's law follows entirely from this mechanical model. In turn, in 1694, Newton provided a more accurate atmospheric refraction correction table by using his model [66,67].

In 1678, Huygens in his Treatise on Light assumed that the medium would reduce the speed of light, and then used the wave theory later known as Huygens principle to obtained Snell's law [68]. Huygens, however, assumed that light is made of longitudinal waves, and therefore he could not explain optical phenomena related to polarization. In the same period, Hooke also claimed that light is a kind of wave [69]. Despite these earlier proposals, due to Newton's great achievements in physics and mathematics, over the next hundred years, although there were only few scientists (notably Euler) supporting wave theory of light, and most people were still convinced by Newton's arguments [69,70].

It was not until 1801, when Young completed the double–slit interference experiment of light, that someone began to study light with wave theory again [69]. It is worth mentioning that Young was presumably the person who first used, and invented, the name "index of refraction" [71]. One of the most outstanding is Fresnel, who assumed that light consists of shear waves, and developed Huygens principle, thus explaining almost all optical phenomena discovered at that time, including reflection, refraction, diffraction, interference, polarization and birefringence [69,72]. In the early stage, Fresnel's view was criticized by Laplace, Poisson, Biot and others [69,70]. In fact, Poisson used Fresnel's wave theory to predict that an opaque circular plate illuminated by monochromatic light would appear as a small bright spot in the shadow center of the light screen, which he thought was an absurd conclusion [69,70]. Not long after, however, Fresnel's friend Arago experimentally confirmed that the bright spot of Poisson's prediction was real [69]. So, it is now generally called Poisson spot [73] or Arago spot [74]. Arago's experiment strongly supports the wave theory of light, which indirectly proves that the speed of light in dense medium is lower than in the vacuum. At the suggestion of Arago, Foucault and Fizeau independently measured the relative velocity of light in air and water in 1850, and the result was that light in water was slower than in air [69,75].

At that point, the controversy over the speed of light in medium seemed to be over, but it quietly reappeared in the form of the A–M controversy. In 1821, at the age of 17, Hamilton began to research geometric optics, and established the Hamiltonian optical theory by mathematical analysis [76]. In this theory, the generalized momentum of light is proportional to the refractive index of the medium [77,78]. Rather obviously, it represents Minkowski momentum. Because Hamiltonian optical theory was too advanced and abstract for his times, it was not discovered and published until 1931, more than 100 years later [76]. In 1905, Poynting proposed the momentum fluid model of light, whose flux is equal to the energy density $U$, based on the theoretical and experimental research of light pressure in a vacuum by Maxwell, Thomson, Lebedew, Nichols, and Hull et al. He then introduced the momentum density of light, $g = U/v$, where $v$ is the speed of light in a vacuum or medium [79]. Since the experiment is the basis of physics, and the light speed in medium as shown in Eq. (2b) is in accordance with the experimental observation, Poynting used Eq. (2b) to give the momentum density of light in medium, $g = nU/c$. It is the same as Minkowski momentum density, but it was proposed earlier [79]. Conversely, if the speed of light in denser medium is assumed to be larger (obviously, this view does not conform to the experiments of Foucault and Fizeau), then the Abraham momentum density $g = U/(nc)$ can be obtained according to Poynting's momentum fluid model [79].

In 1905, Einstein applied Planck's energy quantization hypothesis to light, and put forward the light quantum hypothesis, which well explained the photoelectric effect [80]. Einstein's hypothesis that light can be regarded as a stream of particles is consistent with Newton's view, and thus with the idea that the speed of light in a medium is proportional to the refractive index. According to Planck–Einstein relation $E = hf$ and Poynting's momentum density formula $g = nU/c$, it is easy to obtain the quantum mechanical formula of Minkowski momentum shown in Eq. (1a). However, Einstein did not provide any expression for photon momentum in his original paper. It was not until 1916 that Einstein gave the first expression of photon momentum in vacuum, $p_0 = hf/c = h/\lambda_0$ [81]. The second expression may be extended to medium, and Eq. (1a) is obtained. It is not known whether Einstein in 1916 agreed with such an expansion, but it is certain that Einstein in 1908 must have held a negative attitude towards it, because he and Laub were the first physicists to criticize Minkowski momentum, suggesting the so called Einstein–Laub momentum density [82]. However, in 1918, Einstein changed his view and supported Minkowski momentum density, criticized Abraham momentum density, and declared that the momentum density given by himself and Laub was not tenable [83].

In conclusion, the above discussion illustrates how the A–M controversy is a continuation of the controversy about the speed of light in medium. Both of them can lead to Snell's law, which is the basic law of light propagation, and both of them well explain physical phenomena. With the advancement of technology, both views have been supported by several experiments. Experiments are the ultimate basis for testing theories. This is a concept that was initiated by Galileo [84], the father of modern science and widely recognized, and it is also one of the basic characteristics of modern science. Thus, the opposition of experimental observational evidence is the root cause of the A–M controversy that persists to this day.

## 3 The analog of A–M controversy for an aircraft

### 3.1 Perspective of earth

Assume that the fuel consumption rate (FCR) of an aircraft is proportional to the square of the flight speed, that is, Eq. (5) holds, where $C$ is a constant.

$$\text{speed} = C \times \frac{\text{FCR}}{\text{speed}} \tag{5}$$

In fact, the relationship between fuel consumption rate and speed is very complex and does not strictly conform to the simple relationship shown in Eq. (5). On the other hand, for and engine working close to the cruise speed, Eq. (5) provides a good approximation. Substituting the definition of speed shown in Eq. (6a) into Eq. (5), Eq. (6b) can be obtained.

$$\text{speed} = \frac{\text{distance}}{\text{time}} \tag{6a}$$

$$\text{speed} = C \times \frac{\text{FC}}{\text{distance}} \tag{6b}$$

If the influence of fuel consumption (FC) on aircraft mass is ignored, the momentum of aircraft can be obtained according to Eqs. (6a) and (6b).

$$\text{momentum} = \text{mass} \times \frac{\text{distance}}{\text{time}} \tag{7a}$$

$$\text{momentum} = C \times \text{mass} \times \frac{\text{FC}}{\text{distance}} \tag{7b}$$

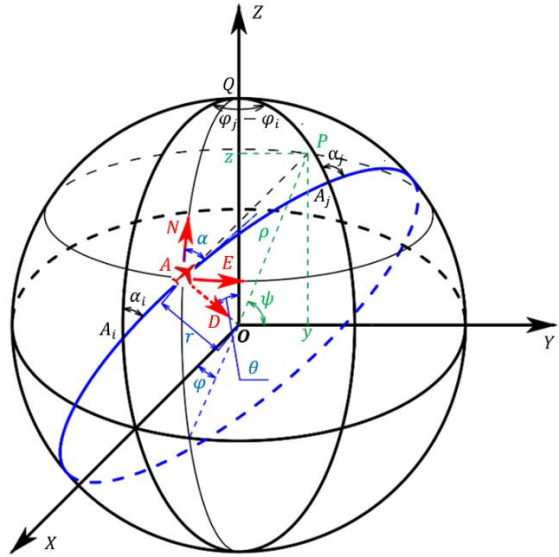

**Fig. 1** Schematic diagram of the great–circle route of the aircraft. The blue curve is the great–circle route of the aircraft, point $A$ is the location of the aircraft, $(r, \theta, \varphi)$ are spherical coordinates of point $A$ in the $OXYZ$ coordinate system, $ANED$ is the navigation coordinate system fixed on the aircraft, $AN$ axis points north, $AE$ axis points east, $AD$ axis points to the center of the earth, $\alpha$ is the heading angle of the aircraft, $P$ is the projection point of the aircraft on the $OYZ$ plane, $(y, z)$ and $(\rho, \psi)$ are the Cartesian coordinates and polar coordinates of point $P$ in $OYZ$ coordinate system respectively, $Q$ is the North Pole.

If the aircraft is to fly along the great–circle route as shown in Fig. 1, the heading angles $\alpha_i$ and $\alpha_j$ of the aircraft at any two points $A_i$ and $A_j$ on the route may be obtained from the basic geometry of the spherical triangle $A_i A_j Q$.

$$\alpha_i = \arctan\left[\frac{\sin\theta_j \sin(\varphi_j - \varphi_i)}{\sin\theta_i \cos\theta_j - \cos\theta_i \sin\theta_j \cos(\varphi_j - \varphi_i)}\right] \tag{8a}$$

$$\alpha_j = \arctan\left[\frac{\sin\theta_i \sin(\varphi_i - \varphi_j)}{\sin\theta_j \cos\theta_i - \cos\theta_j \sin\theta_i \cos(\varphi_i - \varphi_j)}\right] \tag{8b}$$

According to Eqs. (8), the differential equation describing the great circle–route is given by

$$\alpha = \arctan\left[\frac{\sin(\theta + d\theta)\sin(d\varphi)}{\sin\theta \cos(\theta + d\theta) - \cos\theta \sin(\theta + d\theta)\cos(d\varphi)}\right] \tag{9}$$

For an observer from the perspective of the earth, Eqs. (6a) and (6b) are equivalent. The results (7a) and (7b), as obtained from Eqs. (6a) and (6b) are thus the same. If the aircraft flies at a constant speed along the great–circle route, the resultant external force is zero, and the reference frame of the aircraft is an inertial one. The differential equation of motion of aircraft as shown in Eq. (9) can be obtained by using geometric relations. So, from the earth's perspective, there is no controversy about the speed and momentum of aircraft.

### 3.2 Perspective of geosynchronous space station

Let us now assume that there is a geosynchronous space station of earth on the $OX$ axis, where there are two observers $O_1$ and $O_2$. They look at the earth from the space station, just as we look up at the full moon, i.e. a non–rotating disk (not a sphere). It is assumed that $O_1$ and $O_2$ are born in the space station and never left it. So, they had no prior knowledge that the earth is a sphere.

If the aircraft flies at a constant speed along the great–circle at speed $v_0$, the aircraft seen by $O_1$ and $O_2$ is not moving at a constant speed, but moving at a non–uniform speed along a curve. In order to quantitatively describe the motion of the aircraft, $O_1$ and $O_2$ may decide to make remote observations of the physical state of the aircraft. It is mandatory here that $O_1$ and $O_2$ cannot leave the space station and can only observe the aircraft remotely through passive measurement methods, such as optical telescopes.

$O_1$ measures how the position coordinates $(y, z)$ of the aircraft on the plane of the disk with time $t$. Then, through data analysis and mathematical modeling, an empirical formula which describes the trajectory of point $P$ as shown in Eq. (10) may be obtained.

$$\begin{cases} y = r \sin \theta(t) \sin \varphi(t) \\ z = r \cos \theta(t) \end{cases} \quad (10)$$

where, $r$, $\theta$, and $\varphi$ are model parameters introduced by $O_1$ to facilitate modeling. On the other hand, she does not know the geometric meaning of these parameters, only that they could be used to describe the observed data very well. Then, $O_1$ differentiates Eq. (10) with respect to $t$, and obtains the following equations.

$$\begin{cases} \dot{y} = r\dot{\varphi} \sin \theta \cos \varphi + r\dot{\theta} \cos \theta \sin \varphi \\ \dot{z} = -r\dot{\theta} \sin \theta \end{cases} \quad (11)$$

By further analyzing the observed data and introducing new model parameters $v_0$ and $\alpha$, $O_1$ may also establish an empirical formula for $(\dot{\theta}, \dot{\varphi})$ as follows

$$\begin{cases} \dot{\theta} = -\dfrac{v_0 \cos \alpha}{r} \\ \dot{\varphi} = \dfrac{v_0 \sin \alpha}{r \sin \theta} \end{cases} \quad (12)$$

According to physical experience (Newton's law of motion), $O_1$ believes that there must be a lateral force acting on the aircraft, which causes the aircraft's trajectory to bend. She may led to conclude that this is a typical non–inertial reference frame problem. To investigate the nature of this lateral force, $O_1$ decides to calculate the angular momentum $\vec{L}_P$ of the aircraft.

$$\vec{L}_P = \vec{\rho} \times \vec{p}_P = m_0 \vec{\rho} \times \vec{v}_P = m_0 (y\dot{z} - \dot{y}z) \quad (13)$$

where, $\vec{L}_P$ is the vector perpendicular to the plane of the disk, $\vec{\rho}$, $\vec{p}_P$ and $\vec{v}_P$ are the position, momentum and speed vectors of point $P$, respectively.

The expression of $\vec{L}_P$ can be obtained by substituting Eqs. (10)–(12) into Eq. (13).

$$\vec{L}_P = m_0 v_0 r (\sin \varphi \cos \alpha - \cos \theta \cos \varphi \sin \alpha) \quad (14)$$

After $O_1$ substitutes the observation data into Eq. (14), it is found that as long as $m_0$, $v_0$ and $r$ are constants, then $\vec{L}_P$ is a constant. Therefore, $O_1$ considers that the lateral force on the aircraft is similar to the centripetal force in uniform circular motion.

Although $O_1$ concluded that $\vec{L}_P$ is a constant based on the measured data, she does not know the reason behind this phenomenon. In fact, it is easy to prove that $\vec{L}_P$ is a constant by using Eq. (8). Indeed, for any two points $P_i$ and $P_j$ on the aircraft trajectory we have

$$L_{Pj} - L_{Pi} = \left|\vec{L}_{Pj}\right| - \left|\vec{L}_{Pi}\right| = m_0 v_0 r \left(\left|\sin\varphi_j \cos\alpha_j - \cos\theta_j \cos\varphi_j \sin\alpha_j\right| - \left|\sin\varphi_i \cos\alpha_i - \cos\theta_i \cos\varphi_i \sin\alpha_i\right|\right) = 0 \quad (15)$$

The expressions of $\alpha_i$ and $\alpha_j$ as shown in Eq. (8) are substituted into Eq. (15), and then it is easy to prove that Eq. (15) is correct by using symbolic operation tools in MATLAB or Mathematica. The proof process can also be completed by manual derivation, but the process is more complicated. In fact, Eq. (8) and Eq. (14) are equivalent. The former is the geometric expression of aircraft motion from the earth perspective, and the latter is the mechanical expression of aircraft motion from the perspective of space station.

Using the speed $v_P$ of point $P$ and the constant $v_0$, $O_1$ may define the refractive index $n_P$. It is a dimensionless parameter that can be used to characterize the influence of the earth disk on the speed of the aircraft.

$$n_P(\theta, \varphi, \alpha) = \frac{v_0}{v_P} = \frac{v_0}{\sqrt{\dot{y}^2 + \dot{z}^2}} = \frac{1}{\sqrt{(\cos\varphi \sin\alpha - \cos\theta \sin\varphi \cos\alpha)^2 + (\sin\theta \cos\alpha)^2}} \quad (16)$$

$n_P(\theta, \varphi, \alpha)$ can be rewritten to $n_P(y, z, \beta)$ by coordinate transformation. Where $\beta$ is the angle of the speed vector of point $P$ relative to the $OY$ axis, and its expression is shown in Eq. (17).

$$\beta = \arctan \frac{\dot{z}}{\dot{y}} \quad (17)$$

It can be seen from $n_P(y, z, \beta)$ that the influence of the earth disk on motion speed of aircraft is not only related to position coordinates $(y, z)$, but also to the direction $\beta$ of the aircraft. So $O_1$ says that the earth disk can be regarded as an anisotropic medium, and $n_P$ is the refractive index field of that medium. Given the position and motion direction of the aircraft on the earth disk, the speed and the acceleration of the aircraft can be calculated by using $n_P$. According to the definition of $n_P$, $O_1$ draws the following conclusion: the speed $v_P$ and momentum $p_P$ of aircraft are inversely proportional to $n_P$.

$O_2$ may however raise objections to the research methods and conclusions of $O_1$. $O_2$ thinks that it is unscientific for $O_1$ to regard aircraft as a particle because of its internal structure. She thinks that the turbine in the aircraft engine should be taken as the observation object, and the rotating state of the turbine should be connected with the motion state of the aircraft. According to Eq. (6a), $O_2$ may write

$$\frac{\lambda}{T} = f\lambda = v_P \quad (18)$$

where, $f$ is the rotational frequency of the turbine, $\lambda$ is the distance covered by the aircraft on the earth disk in the time $T = 1/f$. If $v_0$ is a constant, then $\lambda_0 = v_0/f$ can be defined according to Eq. (18), and then another formula for $n_P$ can be obtained.

$$n_P = \frac{\lambda_0}{\lambda} \quad (19)$$

In the case of constant engine turbine speed, the fuel consumption (FC) of the aircraft is proportional to the number of revolutions (NOR) of the turbine, which is a well–known quantity. $O_2$ can be obtained from formula (20) accordingly.

$$\frac{FC}{distance} \propto \frac{NOR}{distance} = \frac{NOR}{time} \times \frac{time}{distance} = \frac{f}{v_P} = n_P \frac{f}{v_0} = \frac{n_P}{\lambda_0} \quad (20)$$

It can be seen from Eq. (20) that, under the condition of constant $f$, the fuel consumption of an aircraft within a unit flight distance is inversely proportional to $v_P$, that is, proportional to $n_P$. Using this conclusion, $O_2$ rewrites Eqs. (6b) and (7b) as Eqs. (21) and (22), respectively. It can be seen that they have the same form as Minkowski speed and momentum of photon.

$$v'_P = \frac{H}{m_0 \lambda} = n_P \frac{H}{m_0 \lambda_0} \quad (21)$$

$$p'_P = \frac{H}{\lambda} = n_P \frac{H}{\lambda_0} \tag{22}$$

where, $v'_P$ is the Minkowski speed of the aircraft, $p'_P$ is the Minkowski momentum of the aircraft, $H$ is a constant similar to Planck's constant $h$.

The observer $O_2$, as it does $O_1$, considers the aircraft in a non–inertial frame and subjected to a lateral force, however, she does not agree with description of $O_1$ for lateral forces. $O_2$ thinks that the lateral force on the aircraft should be the derivative of $\vec{p}'_P$ to time $t$, not the derivative of $\vec{p}_P$ to time $t$. These formulas for speed and momentum of aircraft given by $O_1$ and $O_2$ are derived from experimental observations. Due to experiments are the cornerstone of physics and the evidence for testing theories, thus $O_1$ and $O_2$ all think they are right and argue about it.

### 3.3 Discussion

In this example, $O_1$ and $O_2$ may also use the air around the aircraft as an experimentally measurable object. Then, by analyzing the flow state of air, Newton's third law may be used to infer the motion state of aircraft. On the other hand, this indirect method cannot give a clear answer to the A–M controversy for the aircraft. In fact, because the interaction between the aircraft and the atmosphere is very complicated, it will make the controversy even more involved. Similarly, we believe that any experiment that attempts to resolve the A–M controversy by measuring the effect of light on the surface of a liquid and after entering the liquid is not able to lead to a clear answer. On the contrary, taking into account the interaction with the liquid makes the formulation of the controversy more involved, and ultimately confusing.

Physics is an experimental science, and $O_1$'s analysis on aircraft motion conforms to the typical research method in physics. Research on the laws of motion of celestial bodies in the solar system uses this method. Galileo proved through experiments that force is the reason to change the state of motion of matter [84]. Kepler put forward the three laws of planetary motion on the basis of the observation data of Tycho [85]. Newton introduced the three laws describing the motion of matter on the basis of Galileo experiment [86]. Using his own law of motion, Newton used rigorous mathematical derivation to arrive at the law of universal gravitation on the basis of Kepler's law of planetary motion [87]. In the subsequent two hundred years, Newton's law of motion and the law of universal gravitation achieved great success. They can be used to explain almost all mechanical phenomena in the solar system, though no one knows the physical nature of gravity. It was not until Einstein founded general relativity that people knew that gravity is not force [88]. From the perspective of four–dimensional space–time, free falling objects, planets and light all move at a uniform speed along the geodesic (i.e. four–dimensional great–circle route) in the four–dimensional space–time. However, from the perspective of a three–dimensional space, one observes universal gravitation. The situation is analogue to that of $O_1$ and $O_2$ in the perspective of the space station, when they think that an aircraft moving at a constant speed along the great–circle route is subject to a lateral force.

The perspective of the space station is the cause of the A–M controversy of the aircraft. If the motion of the aircraft is described from the perspective of the earth, there is no controversy. From the perspective of the earth, one can use $(\theta, \varphi, \dot{\theta}, \dot{\varphi})$ to describe the motion of an aircraft. Since $\theta$ and $\varphi$ are independent of each other, they are generalized coordinates, and $\dot{\theta}$ and $\dot{\varphi}$ are generalized speeds, which are used to describe the direction of aircraft motion. The set $(\theta, \varphi, \dot{\theta}, \dot{\varphi})$ provides coordinates for a phase space. From the perspective of the space station, one needs to use another set of generalized coordinates and speeds $(y, z, \dot{y}, \dot{z})$ to describe the motion of the aircraft. So, $(\theta, \varphi, \dot{\theta}, \dot{\varphi}) \rightarrow (y, z, \dot{y}, \dot{z})$ is the transformation between two phase spaces and also the transformation between two generalized coordinates. The airplane is not subject to lateral forces in the reference $(\theta, \varphi, \dot{\theta}, \dot{\varphi})$, but it is in $(y, z, \dot{y}, \dot{z})$. It can be seen that the lateral force on the aircraft is similar to the Coriolis and the gravitational force, which can be eliminated by coordinate transformation.

From the perspective of the earth, observers which directly measures the aircraft's motion parameters may be classified into three categories: The first is made by observers which are stationary relative to the earth (denoted by $O_3$,), and in a different position from the aircraft, for example, an air traffic controller. The second class, say $O_4$, are observers which are stationary relative to the earth, and in the same position as the aircraft. Finally, we have observers $O_5$, which are in the aircraft, for example one of the pilots, or a passenger. The maximum distance between $O_3$ and the aircraft is determined by the horizon line of sight, and generally does not exceed 400km. As this value is far less than the average earth radius of 6371km [89], the measurement results recorded by $O_3$ are not really different from those of $O_4$. Since $O_5$ is in the same reference frame as the aircraft, it also represents the observed object. At the same time, since the flying speed of the aircraft is far less than the speed of light, relativistic effects are very small, such that there is almost no difference between the results of measurements made by $O_4$ and those of $O_5$. Therefore, there is no controversy between the three observers about the speed and the momentum of the aircraft.

In summary, we can collectively refer to $O_1$ and $O_2$ as non–local observers, and term their observations non–local measurements. $O_3$ and $O_4$ may be referred to as local observers, and call their observations local measurement. $O_5$ is referred to as the observed. In the perspective of the earth, the observer can make local measurements for the motion state of aircraft, while in the perspective of the space station, the observer can only make non–local measurements. Therefore, it can also be said that the non–local measurement is the cause of the A–M controversy for the aircraft.

## 4 Controversy in the theory of relativity

In the following, we are going to discuss relativity theory from the perspective of non–local observers, and point out how the A–M controversy arises. This proves that the A–M controversy is universal.

### 4.1 Special relativity

Suppose there is a transparent spaceship of length $L$, moving at a constant speed $v$, such that everything that happens in the spaceship can be clearly observed by the people on earth. In the eyes of the astronaut, the time required for her to walk from the bow to the stern at the speed $u$ is $t = L/u$. According to special relativity, earthman believe that the time required for an astronaut to walk from bow to stern is $t'$ and the length of the spacecraft is $L'$, as shown in Eq. (23) [90,91].

$$t' = \frac{t}{\sqrt{1 - \frac{v^2}{c^2}}} \tag{23a}$$

$$L' = L\sqrt{1 - \frac{v^2}{c^2}} \tag{23b}$$

where, $t$ is the clock time in the spaceship observed by the astronaut; $L$ is the length of the spaceship observed by the astronaut; $L'$ is the length of the spaceship observed by an earthman. It's important to note that the $t'$ is not the clock time in the spaceship observed by earthmen, but rather the time passed according to the clock on the earth after the astronaut walked from the bow to the stern. In the eyes of earthman, because the clock time on the spaceship has slowed down, what happens on the spaceship requires longer earth time.

In the eyes of the observer on earth, the speed of astronauts walking is $u'$.

$$u' = \frac{L'}{t'} = u\left(1 - \frac{v^2}{c^2}\right) \tag{24}$$

If a light source at the bow sends a beam of light to the stern, then, in the eyes of the astronaut, the speed of light in the spaceship is $c$. In the eyes of earthman, the speed of light in the spaceship is $c'$.

$$c' = c\left(1 - \frac{v^2}{c^2}\right) \tag{25}$$

In this example, if the object being observed is an astronaut, since $u \ll c$, the astronaut can be regarded as his own local observer. That is to say, astronauts have the dual identities of observed and local observer at the same time. If the observed object is light, the astronaut is a local observer. Therefore, in both cases, the astronaut does not see any contradiction about the discussion of her observations. However, no matter which object is observed, the earthman is a non–local observer. In the eyes of the earthman, all the matter in the spaceship seems to be wrapped in a medium with a refractive index $n_r$, which causes their speed to reduce. We call $n_r$ the relative motion refractive index, and its expression is shown in Eq. (26) [92].

$$n_r = \left(1 - \frac{v^2}{c^2}\right)^{-1} \tag{26}$$

Therefore, in the eyes of a non–local observer, there will be an A–M like controversy over any matter moving in the spaceship, including the light and the walking astronaut.

Let us now employ a more general method to define the relative motion refractive index. If the trajectory $s$ of matter with speed $v$ is taken as the space coordinate, the space–time metric of special relativity can be expressed by Eq. (27) [92].

$$\mathrm{d}\tau^2 = \left(1 - \frac{v^2}{c^2}\right)\mathrm{d}t^2 - \frac{1}{c^2}\left(1 - \frac{v^2}{c^2}\right)^{-1}\mathrm{d}s^2 \tag{27}$$

where, $\mathrm{d}t$, $\mathrm{d}s$, and $\mathrm{d}\tau$ are the infinitesimal time, space trajectory and proper time, respectively.

For Euclidean space, the infinitesimal $\mathrm{d}s$ satisfies the following relations [93].

$$\mathrm{d}s^2 = \mathrm{d}r^2 + r^2\mathrm{d}\theta^2 + r^2\sin^2\theta\mathrm{d}\varphi^2 \tag{28}$$

where, $(r, \theta, \varphi)$ are space spherical coordinates.

For light, $\mathrm{d}\tau = 0$ [92]. So, Eq. (27) can be rewritten as Eq. (29).

$$\left(1 - \frac{v^2}{c^2}\right)\mathrm{d}t^2 - \frac{1}{c^2}\left(1 - \frac{v^2}{c^2}\right)^{-1}\mathrm{d}s^2 = 0 \tag{29}$$

Using the definition of the speed of light in the medium, as shown in Eq. (30) [93], the expression of the relative motion refractive index $n_r$ can be obtained, according to Eq. (29), as in Eq. (31).

$$\frac{\mathrm{d}s}{\mathrm{d}t} = \frac{c}{n} \tag{30}$$

$$n_r = c\frac{\mathrm{d}t}{\mathrm{d}s} = c\sqrt{\frac{\mathrm{d}t^2}{\mathrm{d}s^2}} = \left(1 - \frac{v^2}{c^2}\right)^{-1} \tag{31}$$

### 4.2 General relativity

Let us assume that an interstellar adventurer came to a planet X with a mass much greater than the earth. If she was walking on the surface of the planet X with a flashlight at speed $u$, then the light coming out of her has speed $c$. According to the time dilation and the length contraction in general relativity, an earthman would think that the speed of adventurer is $u/n_g$, and the speed of light from a flashlight is $c/n_g$, where, $n_g$ is gravitational field refractive index, which can be defined by the space–time metric.

If the spherical coordinates $(r, \theta, \varphi)$ are selected as the space coordinates, the space–time metric of the gravitational field can be expressed as [93],

$$\mathrm{d}\tau^2 = g_{tt}\mathrm{d}t^2 + \frac{g_{rr}}{c^2}\mathrm{d}r^2 + \frac{g_{\theta\theta}}{c^2}\mathrm{d}\theta^2 + \frac{g_{\varphi\varphi}}{c^2}\mathrm{d}\varphi^2 + \frac{g_{tr}}{c^2}\mathrm{d}t\mathrm{d}r + \frac{g_{t\theta}}{c^2}\mathrm{d}t\mathrm{d}\theta + \frac{g_{t\varphi}}{c^2}\mathrm{d}t\mathrm{d}\varphi + \frac{g_{r\theta}}{c^2}\mathrm{d}r\mathrm{d}\theta + \frac{g_{r\varphi}}{c^2}\mathrm{d}r\mathrm{d}\varphi + \frac{g_{\theta\varphi}}{c^2}\mathrm{d}\theta\mathrm{d}\varphi \tag{32}$$

where, $g_{\mu\nu}$ is the space–time metric tensor.

Eq. (32) can be written by separating time $t$ and space $s$ variables [93]

$$\mathrm{d}\tau^2 = G_{tt}\mathrm{d}t^2 + \frac{G_{ss}}{c^2}\mathrm{d}s^2 \tag{33}$$

where, $G_{tt}$ and $G_{ss}$ are the metric coefficients associated with $t$ and $s$, respectively. These are determined by

$$G_{tt} = g_{tt} + \frac{g_{tr}}{c}\frac{\mathrm{d}r}{\mathrm{d}t} + \frac{g_{t\theta}}{c}\frac{\mathrm{d}\theta}{\mathrm{d}t} + \frac{g_{t\varphi}}{c}\frac{\mathrm{d}\varphi}{\mathrm{d}t} \tag{34a}$$

$$G_{ss} = \frac{g_{rr}\mathrm{d}s_l^2}{\mathrm{d}s^2} \tag{34b}$$

where, $\mathrm{d}s_l$ is the infinitesimal gravitational field space where the local observer is located. It is determined by

$$\mathrm{d}s_l^2 = \mathrm{d}r^2 + \frac{g_{\theta\theta}}{g_{rr}}\mathrm{d}\theta^2 + \frac{g_{\varphi\varphi}}{g_{rr}}\mathrm{d}\varphi^2 + \frac{g_{r\theta}}{g_{rr}}\mathrm{d}r\mathrm{d}\theta + \frac{g_{r\varphi}}{g_{rr}}\mathrm{d}r\mathrm{d}\varphi + \frac{g_{\theta\varphi}}{g_{rr}}\mathrm{d}\theta\mathrm{d}\varphi \tag{35}$$

For light, $\mathrm{d}\tau = 0$. Eq. (33) can be rewritten into the form of Eq. (36).

$$\frac{\mathrm{d}s^2}{\mathrm{d}t^2} = \frac{c^2}{-\frac{G_{ss}}{G_{tt}}} \qquad (36)$$

By comparing Eq. (36) with Eq. (30), the gravitational field refractive index $n_g$ expressed by space–time metric tensor can be obtained as follows

$$n_g = \sqrt{-\frac{G_{ss}}{G_{tt}}} \qquad (37)$$

Similarly to the spaceship example, the adventurer here is the observed or local observer, while the earthman is the non–local observer. In the eyes of the earthman, all matter on planet X seems to be wrapped in a medium with a refractive index $n_g$, which causes their movement speed to slow down. Therefore, there must be an A–M like controversy in general relativity.

### 4.3 Discussion

One of main claim of the present paper is that there is an A–M controversy in the theory of relativity. In fact, one may wonder why this has been never reported. We argue that this is due to the fact that it is not necessary to treat the three–dimensional space occupied by moving matter (such as a spaceship) as a medium with a refractive index $n_r$, nor is it necessary to treat the three–dimensional gravitational field space (such as the surface of planet X) as a medium with a refractive index $n_g$. For this type of problems, one may use Lorentz transformations [94], which represent a more convenient method to describe transformations between two four–dimensional space–time coordinates. Special relativity requires Lorentz transformations, while general relativity deals with local Lorentz transformations.

The coordinate transformation method can also be used to analyze the aircraft flying at a constant speed along a great–circle route, so it is not different from the theory of relativity in terms of mathematical methods, except that they have different spatial geometry. The aircraft moving at a constant speed along the great circle route is about a two–dimensional curving space embedded in three–dimensional space, whereas special relativity deals with a four–dimensional flat space–time and general relativity involves a four–dimensional curved space–time embedded in five–dimensional space. From the point of view of spatial features, the aircraft moving uniformly along the great–circle route is the same as the matter moving along the four–dimensional space–time geodesic in the gravitational field. Both of them are geometric problems involving N–dimensional curved spaces embedded in N+1–dimensional spaces, where, N is the maximum number of generalized coordinates required to describe physical phenomena. The spatial curvature produces a virtual force, and this force can be eliminated by a coordinate transformation.

Physics is an experimental science, and any theory inconsistent with experiments will be abandoned. The principle of relativity [95], which was first put forward by Galileo, agrees with extensive physical experience. A constant value for the speed of light in vacuum is in full agreement with experimental observations, and the Michelson–Morley experiment [96] is the most powerful evidence. According to the principle of relativity and the assumption of the constant speed of light in vacuum, the Lorentz transformation can be obtained [97]. If two observers conduct the same physical experiment in different places, they can use the same theory to quantitatively describe their results. This is the principle of relativity. For their own experiments, they are local observers, and for each other's experiments, they are non local observers. Non–local observations allow them to find that the other's experimental results are inconsistent with their own. Only through Lorentz transformations one observer may match her results to the other's one.

Since Lorentz transformations provide a very useful mathematical tool, there is certainly no need to use refractive index to deal with relativity. On the other hand, this doesn't mean that the theory of relativity cannot be formulated in terms of a refractive index. As far as we know, no one has ever pointed out the A–M controversy hidden in the theory of relativity. In a recent paper [93], the present authors use the gravitational field refractive index to describe the gravitational deflection and Shapiro time delay of light, which are exactly the same as those obtained from geodesic equation in general relativity. In an earlier paper [98], the authors have calculated the deflection angle and the delay time of light passing over the surface of the sun by taking the gravitational field space of the sun as a special medium, and the results were in good agreement with the previous observations. In addition, a new formula for the perihelion precession has been derived by using the Minkowski speed of the planet, and it has been verified by taking Mercury as an example [99].

There are no physical particles such as atoms/molecules in the four–dimensional space–time of the theory of relativity, which allows people to measure physical phenomena locally as local observers. On the other hand, for the medium composed of atoms and molecules, it seems that Nature does not allow any observer to make local measurements. This is due to the fact that no matter how one measures it, the speed of light in the medium is different from that in vacuum.

Mathematically, the definition of the refractive index of the medium shown in Eq. (30) can be rewritten into the form of the space–time metric in the special relativity shown in Eq. (27), i.e.

$$d\tau^2 = \frac{1}{n}dt^2 - \frac{n}{c^2}ds^2 = 0 \tag{38}$$

Although this form is inconsistent with experimental measurements, it implies that the space–time structure of a non–dispersive transparent medium is the same as the special relativity, and further implies that the intrinsic speed of light in the medium is the same as the speed of light in vacuum. Therefore, we can make the following hypothesis: As long as the photon is not absorbed or reflected by the medium, its intrinsic speed remains unchanged, and its modulus is equal to the speed of light in vacuum.

According to this assumption, Newton's second law of light as shown in Eq. (39) can be obtained

$$F_M = \frac{dp_M}{dt} = \frac{d(nm_0 c)}{dt} = m_0 c \frac{dn}{dt} + nm_0 \frac{dc}{dt} \tag{39}$$

where $F_M$ is the Minkowski force on a photon measured by a non–local observer, here referred to as the Minkowski force of light, and $\frac{dn}{dt}$ represents the rate of change of refractive index along the optical path.

According to the hypothesis of this paper, refraction does not change the intrinsic speed of light. So, for light emitted from air into water, the second term in Eq. (39) is zero, and $F_M$ is only related to the rate of change of the refractive index along the path. The law of refraction can be obtained by using it. We note in passing that this is the same way Descartes and Newton thought about the law of refraction. Concerning reflection, the direction of light changes, but the refractive index of the medium remains unchanged. So, the first term in Eq. (39) is zero, and $F_M$ is only related to the refractive index of the medium and the direction change of light before and after reflection. This can be used to explain the light pressure experiment on a flat mirror immersed in liquid [6–8], as well as the momentum transfer experiment at the water/air interface under total internal reflection [19].

## 5 The force of the sun on light

### 5.1 Problem

According to general relativity, light from distant stars deflects as it passes over the surface of the sun. Einstein first gave the approximate calculation formula shown in Eq. (40) [87]. In the third year after the general relativity was formulated, Eq. (40) [100] was confirmed by experimental observations

$$\Delta\alpha_E \approx \frac{4GM}{r_s c^2} \tag{40}$$

where, $G$ is the gravitational constant, $M$ is the mass of the sun, $r_s$ is the radius of the sun.

If light is regarded as a classical particle, and the law of universal gravitation is applied, the following formula for the deflection angle of light may be obtained [101].

$$\Delta\alpha_N = \frac{2GM}{r_s c^2} \tag{41}$$

It can be seen that the predicted value of Newton's theory is about half of that of general relativity. Why is there such a coincidence? To the best of our knowledge, no one has ever given a simple and rigorous answer to this. In the following we are going to explain this coincidence using Eq. (39) and the gravitational field refractive index of the sun.

### 5.2 Solving

In the eyes of an observer on earth (non–local observer), light passing over the surface of the sun deflects. The speed of light at the point $\vec{r}$ in three–dimensional space is $c/n_g$, but the intrinsic speed of light does not change, that is, the speed of light in vacuum is constant. From Eq. (39), the Minkowski force of the sun gravitational field acting on the photon may be obtained as follows

$$F_M = m_0 c \frac{dn_g}{dt} \tag{42}$$

If the relative permittivity $\varepsilon_r$ and relative permeability $\mu_r$ of the medium change slowly with time (i.e., far less than the speed of light), Maxwell's equation can be simplified to the wave equation [102] as

$$\left(\nabla^2 - \frac{n^2}{c^2}\frac{\partial^2}{\partial t^2}\right)\vec{V}(\vec{r},t) = 0 \tag{43}$$

where, $\nabla^2 = \vec{\nabla} \cdot \vec{\nabla}$ is the Laplace operator, $\vec{\nabla}$ is the Hamiltonian operator, $n = \sqrt{\varepsilon_r \mu_r}$ is the refractive index of the medium, $\vec{V}(\vec{r},t)$ is a physical quantity that fluctuates with time and space position, such as the electric field intensity $\vec{E}$ and/or the magnetic flux density $\vec{B}$.

According to the wave equation we have

$$\vec{\nabla} \cdot \vec{\nabla} = \left(\frac{n}{c}\frac{\partial}{\partial t}\right)\left(\frac{n}{c}\frac{\partial}{\partial t}\right) \tag{44}$$

Let $\vec{\nabla} = \nabla \hat{g}$, where $\hat{g}$ is the unit vector of $g$ along the gradient direction. The scalar form of the gradient operator can be obtained.

$$\nabla = \frac{n}{c}\frac{\partial}{\partial t} \tag{45}$$

where $\nabla$ has only two directions. Positive value indicates $\hat{g}$ direction and negative value indicates $-\hat{g}$ direction.

If $\nabla$ is applied to the static refractive index field $n(\vec{r})$, which is independent of time, Eq. (45) can be rewritten into the total differential form as follows

$$\nabla = \frac{d}{dg} = \frac{n}{c}\frac{d}{dt} \tag{46}$$

where, $dg = \frac{c}{n}dt$. The associated physical meaning is that the optical path along the $\hat{g}$ direction is the smallest. The actual optical path is not necessarily directed as $\hat{g}$, but there is a bias towards $\hat{g}$.

If the sun's angular momentum and net charge are neglected, the gravitational field space–time satisfies the Schwarzschild metric shown in Eq. (47) [99].

$$d\tau^2 = \left(1 - \frac{2GM}{rc^2}\right)dt^2 - \frac{1}{c^2}\left(1 - \frac{2GM}{rc^2}\right)^{-1}dr^2 - \frac{r^2}{c^2}d\Omega^2 \tag{47}$$

where, $(r, \theta, \varphi)$ are the spherical coordinates with origin in the center of mass of the sun, $d\Omega^2 = d\theta^2 + \sin^2\theta d\varphi^2$ is the metric of unit sphere.

By substituting Eq. (47) into Eq. (37), the expression of the gravitational field refractive index of the sun can be obtained.

$$n_g = \left(1 - \frac{2GM}{rc^2}\right)^{-1}\sqrt{1 - \frac{2GM}{rc^2}\frac{r^2 d\Omega^2}{ds^2}} \tag{48}$$

The argument of the square root in Eq. (48) is the difference between two terms. Since the gravitational field of the sun satisfies the weak field condition of $\frac{2GM}{rc^2} \ll 1$, the second term is a negligible high–order small quantity compared to the first one. Therefore, an approximate expression for $n_g$ is given by

$$n_g \approx \left(1 - \frac{2GM}{rc^2}\right)^{-1} \tag{49}$$

Eq. (49) shows that the gravitational field of the sun is spherically symmetric, and $n_g$ decreases with the increase of $r$. So, the gradient direction of the gravitational field refractive index of the sun is $-\vec{r}$. Substituting Eq. (49) into Eq. (46) we have

$$\frac{dn_g}{dr} = \frac{n_g}{c}\frac{dn_g}{dt} \tag{50}$$

whereas substituting Eq. (50) into Eq. (42) leads to

$$F_M = \frac{m_0 c^2}{n_g}\frac{dn_g}{dr} \tag{51}$$

The approximate expression of $F_M$ can be obtained by substituting Eq. (49) into Eq. (51), thus arriving at

$$F_M \approx -\frac{2n_g GM m_0}{r^2} \tag{52}$$

The approximate value $n_g(r_s) \approx 1.00000425$ of the gravitational field refractive index at the surface of the sun can be obtained by using given by: $c = 2.99792458 \times 10^8 \mathrm{ms}^{-1}$ [103], $G = 6.6740831 \times 10^{-11} \mathrm{Nm}^2\mathrm{kg}^{-2}$ [103], $r_s = 6.957 \times 10^5$ km [104], $M = 1.9891 \times 10^{30}$ kg [104]. Since $n_g(r_s)$ is the maximum value of the gravitational field refractive index of the solar system, that is, $n_g \leq n_g(r_s)$, we may approximate its value by 1 and write

$$F_M \approx -\frac{2GM m_0}{r^2} = 2F_N \tag{53}$$

where, $F_N$ is the universal gravitation obtained by treating photons as classical particles with mass $m_0$, here we call it the Newtonian force of light.

It can be seen that the force of the sun on the light passing over its surface is the Minkowski's force, which happens to be about twice as strong as Newtonian force. It is this reason that causes the difference between Eq. (40) and Eq. (41).

### 5.3 Discussion

In addition to gravitational deflection, the sun also has a time delay effect on light passing over its surface. This was discovered by Irwin I. Shapiro in 1964 using general relativity [105]. Several years later, the experimental team led by Shapiro confirmed this phenomenon [106–108]. Therefore, the gravitational time delay is often referred to as the Shapiro time delay. The phenomenon may be rephrased by saying that an observer on earth is a non–local observer and therefore sees light slowing. The speed of light at point $\vec{r}$ in the three–dimensional space becomes $c/n_g$. Obviously, if the Shapiro time delay is studied by mechanical method, it must be assumed that there exists an Abraham force which can slow down the photon. The Abraham momentum of the photon in the gravitational field is $p_A = m_0 c/n_g$, the Abraham force may be thus obtained by taking the time derivative of $p_A$, ie. $F_A \approx 2GM m_0/n_g r^2$. Therefore, in the eyes of a non–local observer, Abraham force is the cause of Shapiro time delay.

It can be seen that the propagation of light in the gravitational field also leads to A–M like controversy. In fact, there is also an A–M controversy for moving celestial bodies in the gravitational field. It's just that the gravitational field refraction index of sun is so close to 1 that it can't be detected in real time by experimental instruments. But long–term observations of planets close to the sun may ultimately find out phenomena related to the A–M controversy. It can be proved that the perihelion precession of planet is caused by its Minkowski speed. Reference [99] explains the perihelion precession of mercury well by using the Minkowski speed. Therefore, the A–M controversy is universal, and it is caused by non–local observations.

According to the above arguments and viewpoint, it is easy to explain the experiment reported by She and Yu et al. [24]. Since the wavelength of light emitted from a nano fiber taper is of the same order of magnitude as the diameter of the fiber, only the time delay effect may occur, but not the refraction effect. So, a non–local observer can only see the Abraham force, which causes the nano fiber taper to bend.

## 6 Discussion and conclusion

In addition to the gravitational deflection of light, the Shapiro time delay and the perihelion precession of mercury, there is also a physical phenomenon related to general relativity, it is gravitational redshift [109]. They are often referred to as the four tests of general relativity. There is only one observer of the first three physical phenomena, and all of them are non–local observers, so the A–M controversy appears. There are two observers for the gravitational redshift, and both of them are local observers. One of the observers may not really exist, but it can exist. For example, the physical parameters of celestial bodies can be inferred from their gravitational redshift [110]. In this example, there is no real local–observer on the surface of the celestial body. But if there were a person who can adapt to the extreme environment of the surface of the celestial body and can get there, then the result of her measurement there must be exactly the same as that of the imaginary local observer. The phenomenon of gravitational redshift may be accounted for by space–time coordinate transformations. This is the principle of relativity. Similarly, the clock problem in general relativity is also a space–time coordinate transformation problem between two local observers. Since there are no non–local observers, there is no A–M controversy in those problems of relativity theory.

For an aircraft moving at a constant speed along the great–circle route and some relativity phenomena, people can measure them either as a local observer or as a non–local observer. If one studies these problems from the perspective of a local observer, the physical problems may be turned into geometric problems, which is very in line with people's thinking habits. It is for this reason that people did not notice the A–M controversy hidden in them. However, for light propagating in a lossless transparent medium, it seems that Nature does not allow people to measure it as a local observer. That's what led to the A–M controversy.

The conclusion here is that nonlocal observation is the cause of the A–M controversy. It has nothing to do with the medium. Even without a medium, there may be an A–M controversy. It is also not the unique property of light, and the non–local observation for any moving matter may lead to the A–M controversy.

**Acknowledgments** The authors would like to express their gratitude to EditSprings (https://www.editsprings.com/) for the expert linguistic services provided.